\documentclass[conference]{IEEEtran}

\ifCLASSINFOpdf
\else
\fi
\usepackage[T1]{fontenc}
\usepackage{lmodern}
\usepackage[utf8]{inputenc}
\usepackage{lmodern}
\usepackage{xspace}
\usepackage{multirow}
\usepackage{amsmath}
\usepackage[font={footnotesize,bf},textformat=period]{caption}
\usepackage[font=footnotesize,textformat=simple]{subcaption}
\usepackage{xspace}
\usepackage{paralist}
\usepackage{tikz}
\usepackage{tikz-qtree}
\usepackage{pgfplots}
\usepackage{pgfplotstable}
\usepackage{etoolbox}
\usepackage[utf8]{inputenc}

\providecommand{\url}[1] {#1}

\usepackage{amssymb}

\usetikzlibrary{patterns,arrows,positioning,shapes,calc,pgfplots.groupplots}
\pgfplotsset{compat=newest, legend style={font=\scriptsize},compat/path replacement=1.5.1}

\definecolor{light-gray}{gray}{0.65}
\definecolor{dark-green}{HTML}{007F00}

\newcommand{\hide}[1] {}

\newif\ifshowdata
\showdatatrue

\newif\ifshownotes
\shownotestrue

\ifshownotes

\newcommand{\mnote}[1] {{$\langle${\textcolor{green!60!black}{Mayur: \textbf{#1}}}$\rangle$}}
\newcommand{\knote}[1] {{$\langle${\textcolor{blue}{K.K.: \textbf{#1}}}$\rangle$}}
\newcommand{\jnote}[1] {{$\langle${\textcolor{green!40!blue}{Jia: \textbf{#1}}}$\rangle$}}
\newcommand{\snote}[1] {{$\langle${\textcolor{red}{Sripriya: \textbf{#1}}}$\rangle$}}
\newcommand{\fnote}[1] {{$\langle${\textcolor{yellow!80!black}{Fu: \textbf{#1}}}$\rangle$}}

\else

\newcommand{\mnote}[1] {}
\newcommand{\knote}[1] {}
\newcommand{\jnote}[1] {}
\newcommand{\mjnote}[1] {}
\newcommand{\fnote}[1] {}
\newcommand{\snote}[1] {}
\fi

\newcommand{\ie}{\emph{i.e.}\xspace}

\newcommand{\Eg}{E.g.\xspace}
\newcommand{\etc}{etc.\xspace}
\newcommand{\etal}{et al.\xspace}

\hide{
\makeatletter
\renewenvironment{itemize}
{\list{$\bullet$}{\leftmargin\z@ \labelwidth\z@
\itemindent-\leftmargin
}}
{\endlist} \makeatother
}

\makeatletter
\renewenvironment{itemize} {
\begin{list}{$\bullet$}
    {
    \setlength{\itemsep}{0pt}
     \setlength{\parsep}{0pt}
     \setlength{\topsep}{0pt}
     \setlength{\partopsep}{0pt}
     \setlength{\leftmargin}{0.9em}
     \setlength{\labelwidth}{0.6em}
     \setlength{\labelsep}{0.4em}
    }
}
{\end{list}}
\hide{
\newcounter{enumerates}

}
\makeatother

\newcounter{numberedSmallTitle}

\begin{document}
%
\title{COPSS-lite: Lightweight ICN Based Pub/Sub for IoT Environments}

\author{\IEEEauthorblockN{Haitao Wang\IEEEauthorrefmark{1},
Sripriya Adhatarao\IEEEauthorrefmark{2},
Mayutan Arumaithurai\IEEEauthorrefmark{2} and
Xiaoming Fu\IEEEauthorrefmark{2}}
\IEEEauthorrefmark{1}Clausthal University of Technology, Germany. Email: haitao.wang@tu-clausthal.de \\
\IEEEauthorrefmark{2}University of G\"ottingen, Germany. Email: \{adhatarao,arumaithurai,fu\}@cs.uni-goettingen.de
}

\maketitle
\begin{abstract}
Information Centric Networking (ICN) is a new networking paradigm that treats content as the first class entity.
It provides content to users without regards to the current location or source of the content.
The publish/subscribe (pub/sub) systems have gained popularity in Internet.
Pub/sub systems dismisses the need for users to request every content of their interest.
Instead, the content is supplied to the interested users (subscribers) as and when it is published.
CCN/NDN are popular ICN proposals widely accepted in the ICN community however, they do not provide an efficient pub/sub mechanism.
COPSS enhances CCN/NDN with an efficient pub/sub capability.

Internet of Things (IoT) is a growing topic of interest in both Academia and Industry.
The current designs for IoT relies on IP.
However, the IoT devices are constrained in their available resources and IP is heavy for their operation.
We observed that IoT's are information centric in nature and hence ICN is a more suitable candidate to support IoT environments.
Although NDN and COPSS work well for the Internet, their current full fledged implementations cannot be used by the resource constrained IoT devices.
CCN-lite is a light weight, inter-operable version of the CCNx protocol for supporting the IoT devices.
However, CCN-lite like its ancestors lacks the support for an efficient pub/sub mechanism.
In this paper, we developed COPSS-lite, an efficient and light weight implementation of pub/sub for IoT. 
COPSS-lite is developed to enhance CCN-lite and also support multi-hop connection by incorporating the famous RPL protocol for low power and lossy networks.
We provide a preliminary evaluation to show proof of operability with real world sensor devices in IoT lab.
Our results show that COPSS-lite is compact, operates on all platforms that support CCN-lite and we observe significant performance benefits with COPSS-lite in IoT environments.

\end{abstract}
\IEEEpeerreviewmaketitle
\vspace{-3mm}
\section{Introduction}
\label{sec-intro}
The prevalent architecture of today's Internet has undergone a massive change in the requirements since its inception.
The initial requirement was mainly to support resource sharing.
This formed the basis for current design of IP which necessitates a binding association of \emph{what} the user needs to \emph{where} it is located.
The resulting communication pattern involves a point-to-point contact between two entities identified with their respective IP addresses.
However, the current use of Internet is mostly dominated with content sharing.
Users are increasingly becoming interested in receiving the content they desire irrespective of its location.
This shift in the use of Internet has resulted in research efforts towards designing an architecture that caters to the present needs.

Information-Centric Networking (ICN) is a new networking paradigm that treats content as the first-class citizen.
In ICN nodes exchange information based on the \emph{Names} of the content instead of the IP addresses of the end points.
This shift from a ``location-based'' network to a ``content-centric'' network entails efficiency for content dissemination, especially when the content may be available at multiple points and also when the provider and/or consumer are mobile.
Many ICN designs incorporate extensive (in-network) caching.
This results in additional performance benefits with the widespread caches in the Internet.
ICN has gained momentum since its outset and is growing rapidly with a highly active community.
Researchers are increasingly proposing new and improved solutions in various areas like routing, forwarding, caching, naming, \etc, with recent interest towards supporting resource constrained devices in the IoT environment.
Popular ICN proposals include Content Centric Networking (CCN\cite{CCN}), Named Data Networking (NDN\cite{NDN}), MobilityFirst\cite{MobilityFirst}, \etc

Internet of Things (IoT) is a growing topic of interest with IoT being the current buzz word in both academia and industry.
IoT mainly refers to a network of devices like electronic appliances, machines, vehicles, Radio Frequency IDentification tags (RFID), step-counter, \etc 
The IoT devices are usually embedded with sensors, actuators, memory and network connectivity.
These devices are mainly used for applications like sensing, monitoring and controlling other applications.
Billions (50 Billion by 2020\cite{Iot_Statistics}) of IoT devices are expected to be connected in the near future to realize many future applications including smart house, smart Industry, smart vehicles and even smart cities.

The current IoT designs are based on the TCP/IP architecture.
However, the devices in IoT networks are usually constrained in their resources with smaller memory, limited computational capacity, limited power supply (battery).
The IoT devices are equipped with constrained Layer2 technologies like IEEE 802.15.4 and Bluetooth LE resulting in much smaller Maximum Transmission Unit (MTU) of 128B compared to the 1280B of MTU used in the Internet.
Additionally, many of the IoT applications operate in remote locations like forests without any human intervention/supervision for longer duration of time.
They incur overhead with IP's point to point connectivity and limited address space moreover security induces additional overhead.
Many such problems are discussed in Shang~\etal\cite{ndnIoT} and Sripriya~\etal\cite{ISI} discuss the need for efficient communication mechanisms in IoT environment.

We observed a strong co-relation between the requirements of IoT applications and the design of ICN.
The devices in an IoT environment are usually interested in the data sensed by other devices in its area of interest.
The requesting device is merely interested in receiving the data irrespective of who is producing it or where it comes from.
The IoT devices are \emph{information centric} and hence we believe the design of ICN is a better fit for IoT.
ICN provides benefits like eliminating the need for point-point connectivity, unbounded namespace, greater availability of data with widespread in-network caching and enhanced security without additional overhead by necessitating each piece of content to be signed by the publisher.
Many IoT devices can also benefit with features like multicast and broadcast supported easily with ICN without incurring additional overhead.

Publish/Subscribe (pub/sub) systems have acquired a substantial portion of the Internet traffic with steady increase in popularity.
In a pub/sub system, the \emph{publishers} produce content and classify them into categories.
Publishers merely produce the content and do not have any knowledge of the receivers called \emph{subscribers} interested in the content.
Similarly, the subscribers do not have any knowledge of the publishers or when the content of their interest will be published.
Subscribers express their interest in receiving content from certain categories of their interest and receive the content as and when it is published.
Interestingly, the communication in IoT environments also mostly follows the pub/sub paradigm where the sensor devices periodically sense and produce the data while an interested node like the Base Station (BS) collects this data.
The sensor devices are only interested in producing the sensed data and are not aware of who it is targeted to or how it is used.
Similarly the BS is not aware of the devices that are producing the data, it is merely interested in the data.
The pub/sub systems have been known to provide greater network scalability, large scale information dissemination and dynamic network topology.
Although, the design of ICN proposals like NDN supports efficient query/response capability, authors in COPSS\cite{COPSS} point out its limitation to support pub/sub efficiently.
They describe the necessary requirements of pub/sub systems: efficiency and scalability.
They propose a content oriented publish/subscribe system called COPSS to enhance NDN with an efficient pub/sub capability.

There are many works that focus on exploiting the features of CCN/NDN and COPSS for Internet and recent Interest towards supporting IoT environments with CCN/NDN.
However, we argue that IoT devices cannot support and does not need the full CCN/NDN stack.
CCN-lite\cite{ccnlite} is a recent effort by the research community towards producing a lightweight implementation for supporting IoT devices.
However, like CCN/NDN, CCN-lite also faces the limitation to support efficient pub/sub features.
In this work, we enhanced the initial design of COPSS to support IoT environments.
We implement the features of COPSS along with the IoT related enhancements with CCN-lite for enabling efficient pub/sub communication in the IoT environments.
Our key contributions include:
\begin{itemize}
\item Analysis of the requirements for IoT environments.
\item COPSS-lite: Enhancements to the initial COPSS design to support IoT environments.
\item Incorporating COPSS-lite design with CCN-lite to provide efficient pub/sub capability to CCN-lite. 
\item A prototype of the COPSS-lite developed in IoT lab\footnote{https://www.iot-lab.info/} using real sensor devices on RIOT\footnote{https://riot-os.org/} OS.
\item Proof of concept with evaluation.
\end{itemize}

\vspace{-2mm}
\section{ICN Architecture}
\label{sec-ICN_architecture}
\hide{\begin{figure}[t!]
\centering
  \includegraphics[width=1.0\linewidth]{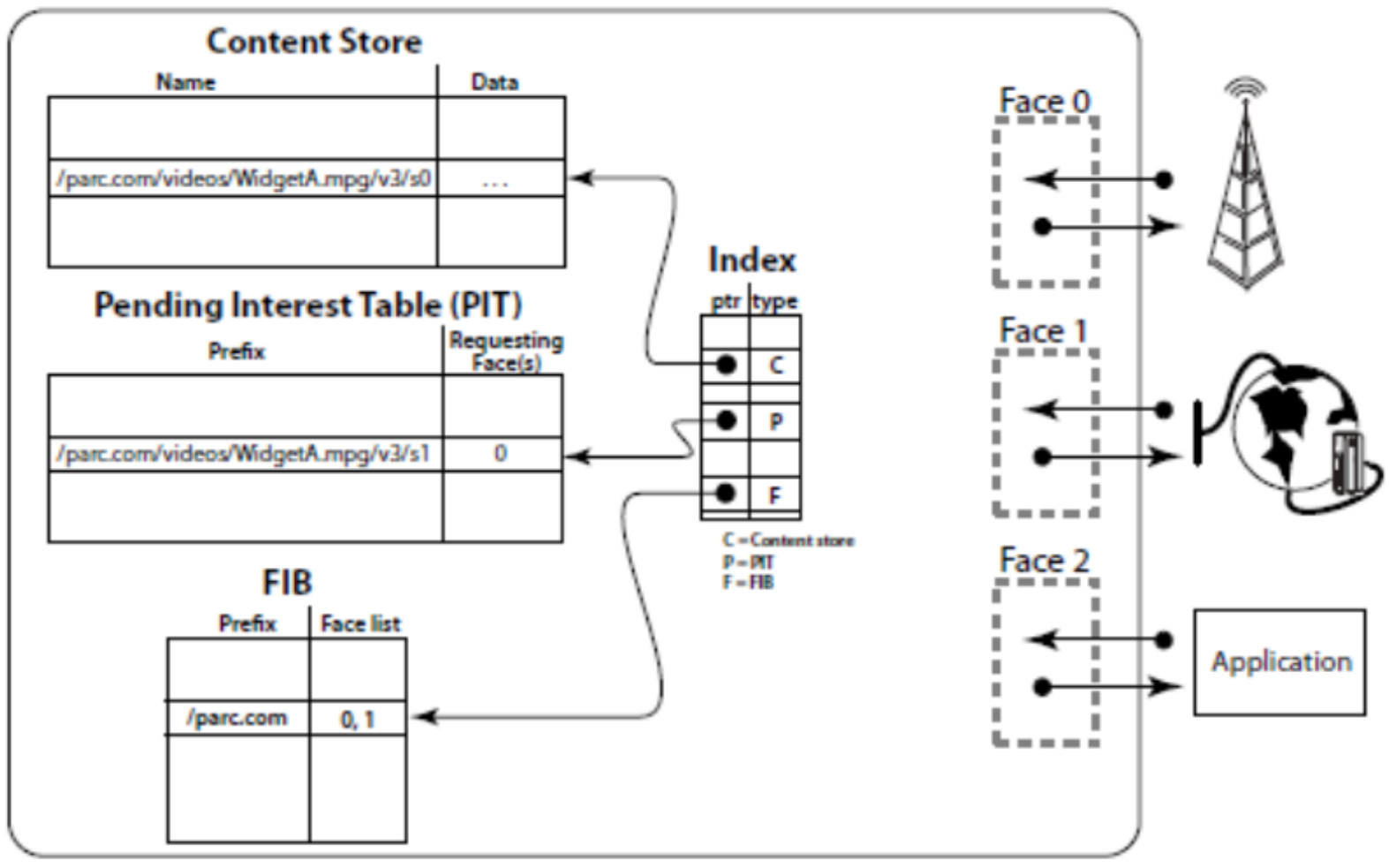}
  \caption{CCN Forwarding Engine Model\cite{NDN}}
  \label{fig-ndn_data_structure}
\end{figure}
}
\hide{
\begin{figure}[t!]
\centering
  \includegraphics[width=0.8\linewidth]{NDN_Architecture_with_COPSS.pdf}
\vspace{-1mm}
  \caption{CCN Forwarding Engine\cite{NDN} with COPSS\cite{COPSS} Subscription Table}
  \label{fig-ndn_copss_data_structure}
\vspace{-6mm}
\end{figure}
}

\hide{\begin{figure}[t!]
\centering
  \includegraphics[width=0.8\linewidth]{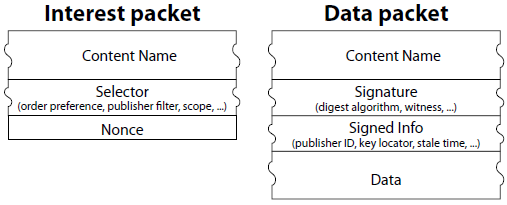}
  \caption{CCN packet types\cite{NDN}}
  \label{fig-ccn_packet_types}
\end{figure}
}

\subsection{NDN}
CCN introduced two new packets: \emph{Interest} (\ie, request) and \emph{Data} (\ie, Content).
NDN uses human-readable and hierarchically structured names as contentnames.
Producers of the content announce prefixes of their sources globally.
In CCN, every router has three data structures: \begin{inparaenum}[\itshape1\upshape)]
\item \emph{Forwarding Information Base (FIB)}: stores content names and their outgoing faces.
\item \emph{Pending Interest Table (PIT)}: stores all the pending Interests and
\item \emph{Content Store (CS)}: caches content received from upstream.
\end{inparaenum}

The communication protocol in NDN begins with a user generating an Interest with the \emph{contentname}.
This Interest is satisfied either by the producer or any intermediate node in the network with a copy of the content.
When a router receives an Interest, it first checks the CS.
If a match is found then the content is sent to the user i.e., the Interest is consumed by the Data.
In case of a CS miss, the PIT is checked to verify if the router has already forwarded an Interest with the same contentname.
If there is a PIT match then the incoming face of the Interest is added to the existing PIT entry.
If there is a PIT miss the FIB performs a Longest Prefix Match (LPM) of the contentname and forwards the Interest and an entry is added to the PIT.
When a router receives a Data packet, the CS is checked, if a matching entry is found, the Data packet is discarded.
In cases of a CS miss, PIT is checked, if an entry is found in PIT then the Data packet is forwarded on all the faces recorded for that entry.
The PIT entry is removed and CS stores the content. In case of PIT miss, the Data packet is unsolicited and hence discarded.
\hide{\begin{figure}[t!]
\centering
  \includegraphics[width=1.1\linewidth]{CCN_lite_Data_structure.png}
  \caption[] {CCN-lite Internal Data structure\footnotemark}
  \label{fig-ccn_lite_data_structure}
\end{figure}
\footnotetext{Source: https://github.com/cn-uofbasel/ccn-lite/blob/master/doc/internal/ccnl-datastruct.pdf}
}
\vspace{-2mm}
\subsection{CCN-lite}
The CCN-lite project is a step towards supporting resource constrained devices to run ICN protocols.
CCN-lite is a light weight implementation of the CCN protocol.
The rationale for developing CCN-lite was by 2011, PARC's CCNx routers had acquired wide acceptance and was growing in popularity.
CCN-lite mainly supports educational and experimental needs without a full fledged implementation of all CCN features.

The CCN-lite is an inter-operable implementation of the CCNx and NDN protocol that supports CCN, NDN, Named Function Networking (NFN\cite{NFN}) and IoT.
CCN-lite provides the core CCNx functionalities with less than 2000 lines of code in \emph{C} language and has a low memory footprint.
It supports multiple platforms like Linux, Android, Arduino, RFduino and OMNET++.
Further, RIOT the free, open source operating system designed specially for IoT has incorporated CCN-lite in its implementation to allow various IoT devices to run ICN protocols.


\vspace{-2mm}
\subsection{COPSS}
COPSS adds a Subscription Table (ST) to the initial CCN forwarding engine.
Every COPSS router maintains the ST for enabling pub/sub communication in a distributed and aggregated manner.
COPSS also introduces two new packets: \emph{Subscribe} and \emph{Publish}.
COPSS uses Content Descriptors (CD) which are some features of the content like keywords, date of publication, identity of the publisher, \etc
The Subscribe packet is used by the users to subscribe to a certain CD while the Publish packet is used by the publisher to publish content for a single or group of CD's \ie a piece of content can be associated with multiple CD's.
Each CD is assigned to a respective Rendezvous Point (RP) node in the network.
Subscribers subscribe to the content of their interest by subscribing to the associated CD's, resulting in a join.
When a router receives a subscription request, it adds an entry in its ST with the CD and the incoming face of the request and forwards the request upstream.
The subscription request is forwarded towards RP with each intermediate router adding an entry for the subscription. 
When the publisher generates any content, (s)he associates the content with the CD's.
Whenever the content is published, it travels along the multicast tree towards the responsible RP.
When a router receives a publication packet, it checks the ST for any one of the CD's in the publication.
If a match is found then the router forwards the publication along the matched interfaces.
However, only a single copy of the publication is forwarded on each interface, irrespective of the number of subscribers downstream. This avoids, unnecessary traffic and duplicate/multiple copies delivered to the subscribers of more than one CD in the publication.


\vspace{-3mm}
\section{IoT Requirements}
\label{sec-requirements}
IoT environment allows numerous physical devices in the sensor networks to integrate with the Internet.
Such an integration of the physical world into the software based Internet enables the possibility to expedite automation in many areas. 
The design of IoT environments varies greatly compared to the design of the Internet be it the traditional IP or the ICN based network.
The nature of the devices operated in the IoT environment is also very different and especially constrained in resources like computational capacity, memory, power, \etc
Even though, the IP based designs are emerging for supporting the IP based IoT networks, we argue that IoT environments are content centric in nature and hence ICN based solutions are more suitable for IoT networks.
We identify the following important requirements for the IoT environments.

\textbf{Communication Protocol:} In an IoT environment, since most of the IoT application tend to operate in a publish/subscribe fashion the pull based mechanism to retrieve the data can turn out to be a huge disadvantage especially in terms of the network load, processing capacity and power consumption. It is desired for the underlying network to provide an efficient platform for pub/sub to enable these devices to operate optimally. 

\textbf{Timeliness:} Many IoT applications are sensitive to time. \Eg emergency applications, environmental monitoring, disaster notifications, Industrial alarming systems, \etc Hence the underlying communication protocol should be reactive and ensure timely delivery of the messages in the IoT environment.

\textbf{Power Consumption:} Many IoT devices operate with limited power supply, mostly a battery. Some devices are expected to operate in remote locations without any human intervention/supervision for longer periods of time. Hence it is crucial for these devices to save power to increase their operational life time. Many devices also tend to sleep during most of their life to save power. The underlying network protocol should support such behaviour and efficiently form the network with available devices and ensure connectivity at all times.

\vspace{-3mm}
\section{COPSS-lite Architecture}
\label{sec-copsslite_architecture}
In this work, we designed COPSS-lite mainly to support the IoT environments.
The IoT environments are usually composed of hybrid devices that mostly sense and produce data of interest for applications that intent to gather, control or monitor the data.
The IoT devices tend to operate in a pub/sub paradigm where we can identify the various sensor devices that periodically sense some data take the role of the publishers.
While the nodes, usually a Base Station (Sink) interested in receiving this data behaves like a subscriber.
A reverse scenario is where a Base Station Controller (BSC) sends some messages/instructions to control the operation of other devices in the network.
In this scenario, the BSC behaves like a publisher while the devices that receive control messages and act on it behave like subscribers.
The pub/sub models are very popular in the current IP based Internet.
However, the current pub/sub systems in Internet are supported mainly via the client side applications and servers to deliver the published content to subscribers.
This entails additional responsibility/burden for the IP based solutions.
In\cite{Multicache,Corona} authors show that IP based solutions tend to waste network resources.
Hence a Content-Oriented Pub/Sub System (COPSS) is necessary to enhance the initial design of CCN by integrating it with an efficient and scalable pub/sub capability not only for the devices in Internet but also for IoT environments (COPSS-lite).


\begin{table}[t]
\centering
\caption{Code Size Estimations}
\vspace{-2mm}
\label{table-code_base}
\begin{tabular}{|l|l|}
\hline
 RIOT& 51540 KB\\ \hline
 CCN-lite&18212 KB\\ \hline
COPSS-lite&192 kB\\ \hline
\end{tabular}
\end{table}
\vspace{-3mm}

We discern that most of the IoT devices are heavily constrained in their available resources compared to their Internet counterparts.
COPSS-lite architecture is designed to add the IoT related enhancements to the original COPSS architecture.
COPSS-lite adds an efficient pub/sub model to the CCN-lite protocol for IoT environments. 
COPSS-lite aims to add several additional features that can support the resource constrained IoT devices.
One of the main requirement we identified is the timely delivery of the information without cutting down the operational lifetime of the devices.
The pub/sub based nature of COPSS-lite reduces the latency of information retrieval (see \S\ref{sec-evaluation}) compared to the pull based mechanism in CCN-lite.
To support the other important requirement of reducing the power consumption we developed COPSS-lite as a compact version of COPSS with CCN-lite. 
CCN-lite is currently implemented using UDP or TCP encapsulation for exchanging the CCN-lite packets.
Our COPSS-lite implementation is written purely in C and is backward compatible with the CCN-lite code.
COPSS-lite adds only 334 lines of code to the original CCN-lite code base\footnote{We are preparing the code for a public release for EUCNC 2017}, the details are shown in the Table \ref{table-code_base}.
We understand that having a smaller code base is necessary for the IoT devices and can greatly benefit not only in reducing the consumption of power but also in terms of memory and reduced computational overhead.

Since many IoT devices show an inclination towards sleeping, it raises a concern on the effective placement of the RP.
Since RP is crucial to enable the pub/sub communication in COPSS-lite, we believe the need to identify the responsible node is imminent.
However, there is no one best answer to this question as the nature of sensor networks varies greatly from one application to another.
We believe that pointing out the need is essential.
We suggest to assign RP to nodes that are either awake all the time (enabled with continuous power supply) or is awake when any other node is awake in the network. The selection of the RP node in an IoT network should ensure that RP is reachable by any node that is awake in the network.

Further, an effective pub/sub solution should also address the necessity to limit data loss, especially in a network where most of the nodes are sleeping and hence not operational at all times.
We propose to utilize a powerful node like the Gateway in\cite{ISI} or to equip the sensor network with a controller node similar to SDN.
The controller keeps all the information about the devices in its sensor network.
The controller node is aware of the sleeping pattern of the devices and hence can efficiently create ad-hoc networks.
Such a network will allow reactive forwarding of the content effectively utilizing the network and ensuring maximum packet transmission with minimal loss.
The controller also stores the state information of all the nodes like their subscription table, and computes routing tables (FIB) for the current duration.
The controller can also store the publications while the nodes are sleeping and propagate it when the nodes wake-up.
Every node has a logical one-hop connection to the controller.
Whenever a node wakes up it contacts the controller and retrieves its state information and joins the network.
Once the network is created, the messages can flow seamlessly.

A two-step communication process in COPSS allows the publisher to exercise control over the access and policy of the published contents. In a two-step communication process, the publishers generate a snippet of the original content with the associated CD's and send it to subscribers. Interested subscribers can then use these CD's to subscribe to the full-length original content. The publishers then publish the original content which is forwarded to the interested subscribers as explained previously.
COPSS-lite also utilizes the two-step communication process for the IoT environment.
Whenever a new device join the network, it can send a snippet of the data that it wishes to publish along with the CD's to the controller.
The controller will disseminate this publication advertisement in the network and the interested users can subscribe to it.
The controller in our design ensures minimum overhead on the constrained devices, minimal loss of publications/subscriptions and maximum utilization of the network.





\vspace{-4mm}
\section{Implementation}
\label{sec-implementation}

\subsection{Challenges}
Even though our choice for OS (RIOT) and CCN (CCN-lite) implementation was clear, developing COPSS-lite from COPSS was not straight forward.
We faced several challenges mainly:
\begin{itemize}
\item CCN-lite works with RIOT on layer 2, whereas COPSS was implemented to work with transport layer with UDP.

\item IoT networks are unstable and need multi-hop routing support without any layer 3 devices.

\item CCN-lite has reduced some original features of NDN that are needed in COPSS like registering prefixes, \etc
\end{itemize}


\subsection{Design}
\begin{figure}[t]
\centering
  \includegraphics[width=0.8\linewidth]{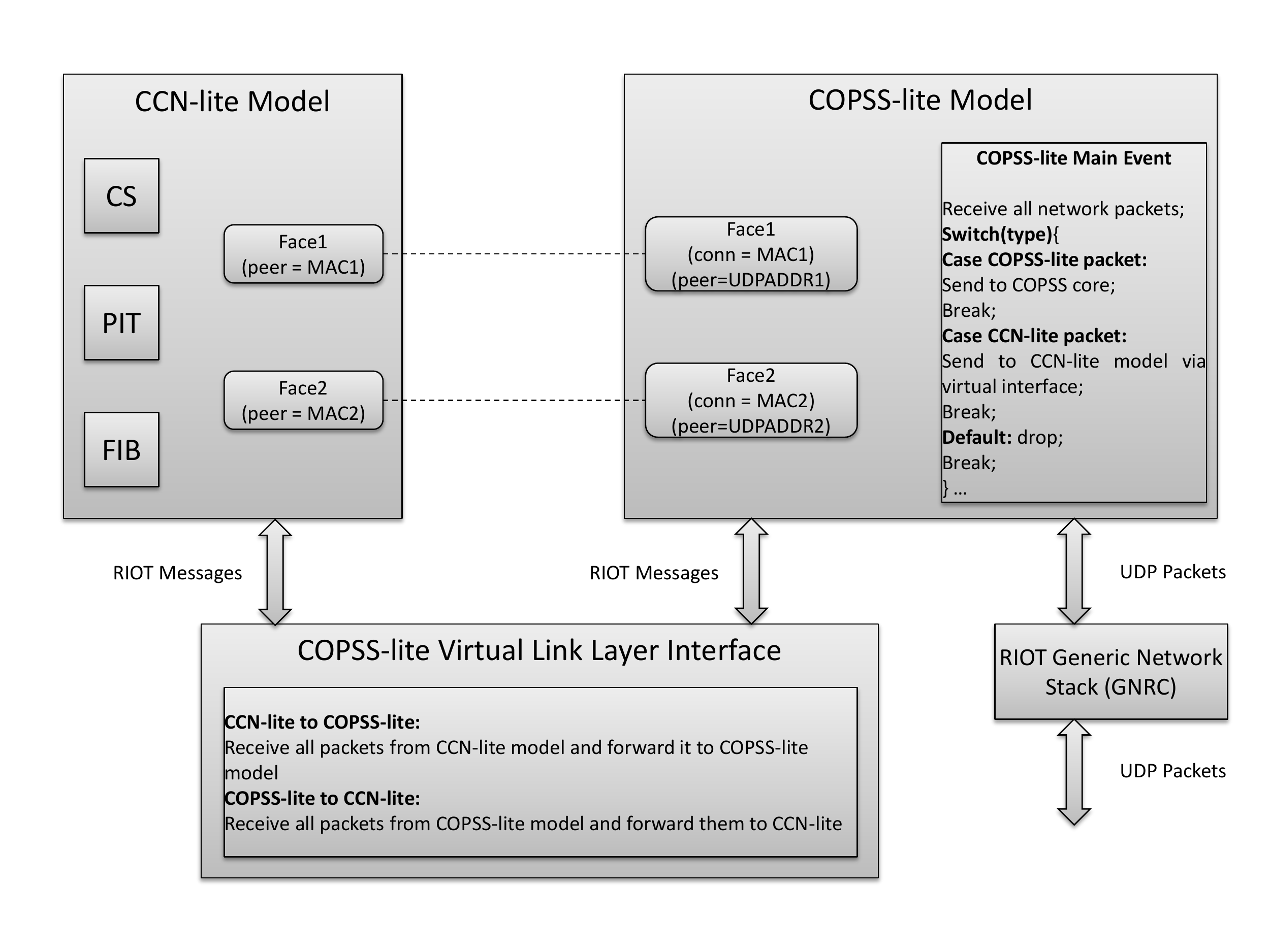}
\vspace{-6mm}
  \caption{Components and Message Exchange}
  \label{fig-structure}
 \vspace{-6mm}
\end{figure}
\vspace{-2mm}
We developed COPSS-lite with the latest versions of RIOT, CCN-lite and the GCC compiler, the details are shown in Table \ref{table-hardware_specifications}(a).
In COPSS-lite we use the Type-Length-Value format for the subscription and publication packets to ensure compatibility and consistency with CCN-lite implementation on RIOT.
We also analyzed the required packet structure for the COPSS packets and redefined the serialization dictionary to meet the requirements.
Further, we developed a design that loosely couples the interaction of CCN-lite and COPSS-lite model with the underlying RIOT system.

Figure \ref{fig-structure} shows the components and the message exchange between CCN-lite, RIOT and COPSS-lite.
We add an intermediate layer \emph{COPSS-lite Virtual Interface} to build a bridge between the CCN-lite and COPSS-lite models.
The CCN-lite model refers to the current CCN-lite implementation in RIOT while the COPSS-lite model refers to the implemented COPSS-lite logic.
The COPSS-lite virtual link layer interface is the adapter between the CCN-lite and COPSS-lite.
The CCN-lite works on the link layer in RIOT and the packets are transmitted using the MAC addresses.
In the COPSS-lite component, each time a user creates a face, the COPSS-lite creates an associating face in CCN-lite and assigns a unique virtual MAC address to it.
The COPSS-lite virtual link layer interface is created to replace the default RIOT link layer interface for CCN-lite so that the CCN-lite packets will be forwarded to the COPSS-lite virtual interface instead.
COPSS-lite module is the main pub/sub component added by the COPSS-lite implementation.
The COPSS-lite model has one main thread and two sub modules: COPSS-lite core and COPSS-lite forwarder.
The Main thread receives all the packets and determines which submodule should process the packet.
The COPSS-lite module deals with the subscribe and publish packets while the COPSS-lite forwarder forwards the CCN-lite Interest and Data packets to the CCN-lite model via the COPSS-lite virtual link layer interface.

\vspace{-2mm}
\section{Evaluation}
\label{sec-evaluation}
\vspace{5mm}
\vspace{-5mm}
\begin{table}
\caption{Development Platform and Hardware Details}
\vspace{-5mm}
\label{table-hardware_specifications}
\begin{minipage}[t]{0.38\linewidth}\centering
  \subcaption{Development Platform}
  \vspace{-2mm}
  \setlength\tabcolsep{1.5pt}
\begin{tabular}{|l|l|}
\hline
OS&RIOT,Linux  \\ \hline
ICN&CCN-lite0.3.0  \\ \hline
Compiler &GCC4.8.4  \\ \hline
Hardware &M3,A8  \\ \hline
\end{tabular}
\end{minipage}
\begin{minipage}[t]{0.45\linewidth}\centering
  \subcaption{M3 and A8 Node}
  \vspace{-2mm}
  \setlength\tabcolsep{1.5pt}
\begin{tabular}{|l|l|}
\hline
 MCU&ARM CORTEX M3, 32-bits  \\ \hline
 Radio&802.15.4  \\ \hline
Power&3.7VLiPo battery  \\ \hline
OS & M3\{FreeRTOS, Contiki, RIOT\} A8\{Linux\}  \\ \hline
ICN&CCN-lite  \\ \hline
\end{tabular}
\end{minipage}
\end{table}

\begin{figure}[t!]
\centering
  \includegraphics[width=0.7\linewidth]{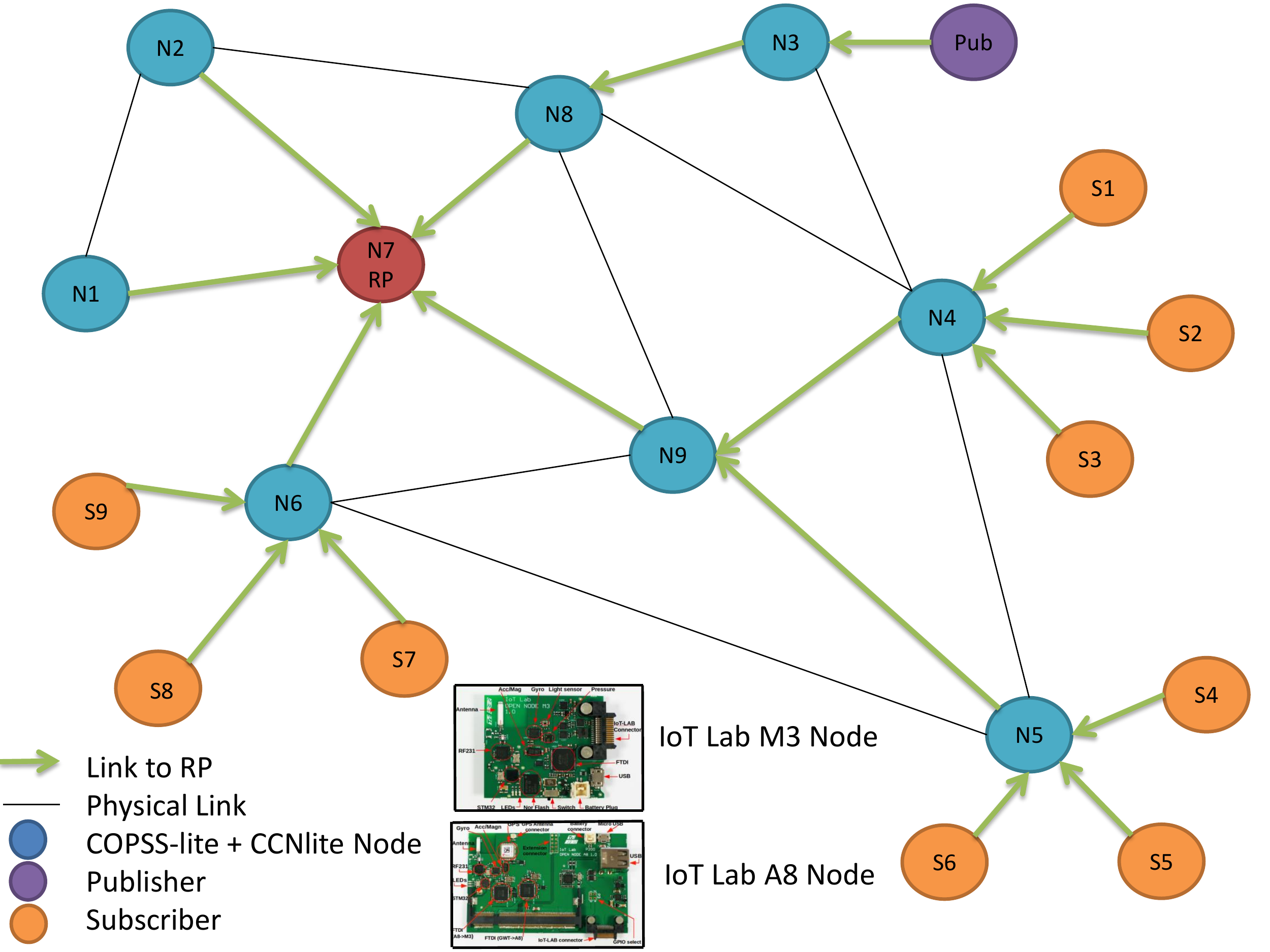}
\vspace{-2mm}
  \caption{Topology}
  \label{fig-topology}
\vspace{-6mm}
\end{figure}
\vspace{-1mm}

\begin{figure}[t!]
\centering
  \includegraphics[width=1.0\linewidth]{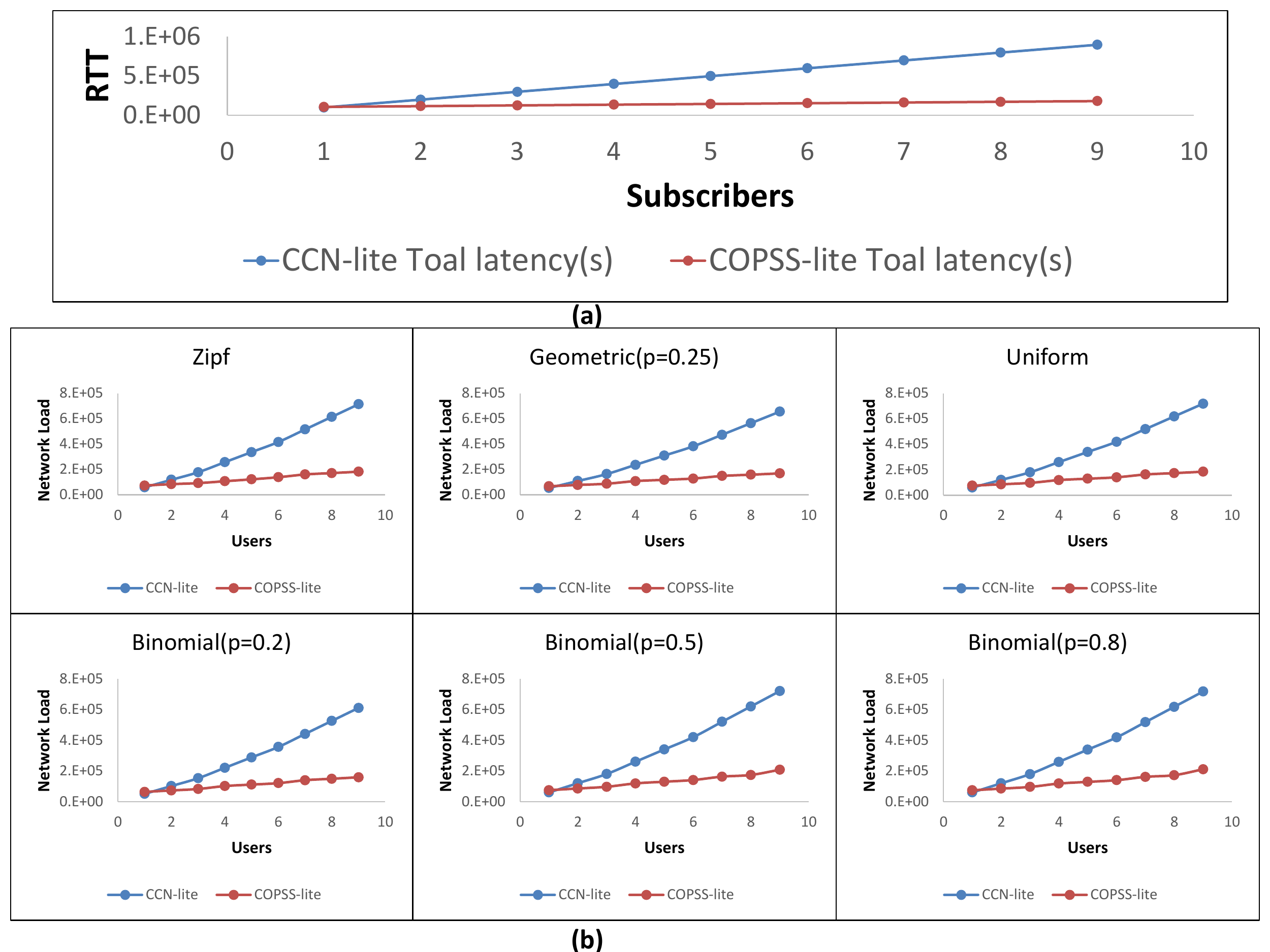}
\vspace{-6mm}
  \caption{(a) Latency and (b) Network Load with different traffic patterns}
  \label{fig-networkload_latency}
\vspace{-8mm}
\end{figure}

To provide a proof of concept and to show that COPSS-lite is operational with real sensor devices we used the IoT Lab to perform our tests.
IoT lab provides a large scale infrastructure facility with real sensor devices for experimentation.
The IoT lab offers three development boards for experimentation: WSN430, M3 and A8 nodes.
We used the M3 and A8 nodes to run our evaluations as they offer a 32-bit system and support CCN-lite (WSN430 is 16-bit).
The development platform and hardware specifications are shown in Table \ref{table-hardware_specifications}(b).
Figure \ref{fig-topology} shows the multi-hop topology used for evaluation.
We measure the aggregated network load and latency experienced by the users with the pub/sub in COPSS-lite and the query/response in CCN-lite.
We comprehend that there are numerous different IoT applications, each with their own peculiar traffic pattern.
Hence we determined that it would be reasonable to test with different distributions that mimic the traffic pattern in the IoT environments.
We used the popular distributions of Zipf, Geometric, Uniform and Binomial to represent the various traffic patterns in the IoT environments for our evaluation.
The Zipf distribution shows that the most popular content with the highest rank is published more often than the content with the lower ranks.
In the Geometric distribution the probability sequence for publishing a piece of content follows a geometric sequence.
The Uniform distribution shows a traffic pattern with constant probability for publishing the content.
While the Binomial distribution shows the number of success in \emph{n} independent trails with the probability of \emph{p}.
We generated a small dataset with 10 different kinds of content to mimic the data generated by the sensor devices.
As shown in the topology, we configured nine nodes to run the CCN-lite and COPSS-lite software with the RIOT OS.
We increased the number of subscribers starting form 1 to 9 with a single producer publishing the content with different distributions.

To test the effectiveness, we published the data with Zipf, Geometric (p=0.25), Uniform and Binomial (p=0.2,0.5,0.8) distributions.
The figure \ref{fig-networkload_latency} shows the resulting user experienced latency in (a) and network load in (b).
We see from the results that network load and the latency experienced by the user is significantly less in IoT environment with the pub/sub based COPSS-lite compared to the query/response based CCN-lite.
Hence we believe the pub/sub based IoT devices can operate efficiently and optimally with their constrained resources in the ICN environment using COPSS-lite.

\vspace{-2mm}
\section{Conclusion}
\label{sec-conclusion}
In this paper, we assessed the changing requirements of Internet users and the necessity for adapting the initial design of IP to current needs.
We then discussed the ICN paradigm and its benefits followed by a discussion of the popular ICN proposals like CCN/NDN.
We focussed especially on the IoT environments highlighting their imminent growth in the near future.
We discussed the potential problems that resource constrained IoT devices face with IP.
We also emphasized on the inherent content centric nature of the IoT environments and suggested ICN for IoT.
We then discussed CCN-lite which provides just enough features to run the ICN protocol in IoT environments.
However, we saw that CCN-lite like its predecessors does not provide an efficient pub/sub mechanism for the IoT environments.
Moreover, we observed the preferred communication protocol in IoT environments is pub/sub.
Hence, we developed COPSS-lite, a light-weight, inter-operable version of pub/sub for IoT.
We developed COPSS-lite and incorporated it with CCN-lite and COPSS-lite can be used on all platforms that support CCN-lite.
With real world sensor devices we provided a proof of operability through evaluation in IoT Lab and show that IoT devices can greatly benefit with COPSS-lite.
\hide{

\snote{
Submission site:https://edas.info/newPaper.php?c=23146
Upload copyriteform}

}

\vspace{-2mm}
\bibliographystyle{IEEEtran}
\bibliography{ccnbib}

\end{document}